\documentclass[conference]{IEEEtran}
\usepackage[utf8]{inputenc}
\usepackage{cite}
\usepackage{amsmath,amssymb,amsfonts, amsthm}
\usepackage{dsfont}
\usepackage{mathtools}
\usepackage[flushleft]{threeparttable}
\usepackage{enumitem}

\usepackage[noend]{algorithmic}
\usepackage[ruled,linesnumbered]{algorithm2e}
\usepackage{scrextend}

\usepackage[subrefformat=parens,labelformat=parens,caption=false,font=footnotesize]{subfig}

\usepackage{tikz}
\usetikzlibrary{shapes,arrows}
\usetikzlibrary{calc}

\newtheorem{remark}{Remark} 

\usepackage{siunitx}

\newcommand\given[1][]{\:#1\vert\:}
\newcommand{\wrt}{\textit{w.r.t.~}}

\newcommand{\ie}{\textit{i.e.~}}

\newcommand{\eg}{\textit{e.g.~}}
\newcommand{\egn}{\textit{e.g.}}

\def\logg{\mathrm{log}}
\def\card{\mathrm{card}}

\def\GHz{~\mathrm{GHz}}
\def\MHz{~\mathrm{MHz}}
\def\Hz{~\mathrm{Hz}}
\def\Mbps{~\mathrm{Mbps}}
\def\m{~\mathrm{m}}

\def\dB{~\mathrm{dB}}
\def\dBi{~\mathrm{dBi}}
\def\dBm{~\mathrm{dBm}}

\newcommand{\mathrmbold}[1]{\boldsymbol{\mathrm{#1}}}
\def\params{\mathrmbold{w}}
\def\tablescale{1}

\newcommand{\titleheader}{This work has been accepted for publication in 2023 IEEE International Conference on Communications (ICC): SAC Aerial Communications Track}

\makeatletter
\def\ps@headings{%
\def\@oddhead{\mbox{}\scriptsize \titleheader}%
\def\@oddfoot{\scriptsize \@date \hfil }%
}

\def\ps@IEEEtitlepagestyle{%
\def\@oddhead{\mbox{}\scriptsize \titleheader \rightmark \hfil }%
}
\makeatother

\begin{document}
\RestyleAlgo{ruled}

\title{Dual-Attention Deep Reinforcement Learning for \\
Multi-MAP 3D Trajectory Optimization \\
in Dynamic 5G Networks }

\author{\IEEEauthorblockN{Esteban Catté, Mohamed Sana and Mickael Maman}

\IEEEauthorblockA{ CEA-Leti, Universite Grenoble Alpes, F-38000 Grenoble, France \\
\{esteban.catte, mohamed.sana, mickael.maman\}@cea.fr\\}
}

\maketitle

\maketitle
\IEEEpeerreviewmaketitle

\begin{abstract}

5G and beyond networks need to provide dynamic and efficient infrastructure management to better adapt to time-varying user behaviors (e.g., user mobility, interference, user traffic and evolution of the network topology). In this paper, we propose to manage the trajectory of Mobile Access Points (MAPs) under all these dynamic constraints with reduced complexity. We first formulate the placement problem to manage MAPs over time. Our solution addresses time-varying user traffic and user mobility through a Multi-Agent Deep Reinforcement Learning (MADRL). To achieve real-time behavior, the proposed solution learns to perform distributed assignment of MAP-user positions and schedules the MAP path among all users without centralized user's clustering feedback. Our solution exploits a dual-attention MADRL model via proximal policy optimization to dynamically move MAPs in 3D. The dual-attention takes into account information from both users and MAPs. The cooperation mechanism of our solution allows to manage different scenarios, without a priory information and without re-training, which significantly reduces complexity.

\end{abstract}
\section{Introduction}

In 5G and beyond networks, it is important to ensure that equal opportunity is offered to users regardless of their location and mobility with a dynamic and efficient management of the infrastructure. This flexible infrastructure can be implemented as a service \cite{Maman2022b} by using Mobile Access Points (MAPs) and will better adapt to time-varying user behavior.
In recent years, the deployment of these MAPs has been widely studied to improve the flexibility of self-managed infrastructures \cite{Catte2022}. However, their management is still poorly studied in highly dynamic networks, taking into account \textit{i)} user mobility, \textit{ii)} interference, \textit{iii)} time-varying user traffic and \textit{iv)} changing scenarios. This paradigm needs to balance the real-time placement of MAPs by tracking the evolution of each user's state to improve the experience with long-term behavior to optimize network resources without adding much complexity to network operation. Our objective is to manage the trajectory of MAPs under all these dynamic constraints with reduced complexity.

In the literature, the MAP management firstly followed centralized and combinatorial approaches.
Authors in \cite{Alsharoa2018} presented a mixed integer linear problem solved iteratively taking into account user's mobility. 
To include user's demand, authors in \cite{Wu2018} designed a successive convex optimization algorithm. It maximizes the average throughput and place MAPs over time.
Authors in \cite{Pan2022} tackled the co-channel interference and MAPs completion time while maximizing average throughput using a particle swarm optimization. These approaches finds optimal solutions for simplified models but fails to handle all constraints at the same time. These methods resolve successive deployment problem over time to create a continuous behavior and adapt to different scenarios with high complexity.

By considering each MAP as an agent, Reinforcement Learning (RL) and Deep Reinforcement Learning (DRL) approaches reduce the complex solution to a single Q-function method for creating complex behaviors. 
The authors of \cite{Rozhina2018} address 3D placement with moving users for one MAP. Then, the approach was extended to multiple MAPs in \cite{Liu2019} but they consider a short time scale execution via Q-learning.
The authors of \cite{Saxena2019} designed a DRL model where each agent acts according to heterogeneous users traffic distribution. Authors in \cite{Zhang2021} proposed a DRL model with filtered actions to optimize the sum-rate of moving users while considering interference. The authors of \cite{Ji2022} and \cite{Ding2020} proposed a dual-clip Proximal Policy Optimization (PPO) algorithm and an actor-critic DRL framework, respectively, to optimize MAP placement and user association at the same time.
These solutions deal with a small evolution of scenarios and not all dynamic constraints at the same time, as the function becomes non-trivial for large parameter sets. To handle time-varying scenarios, they need to be re-trained, which makes it difficult to apply them to highly dynamic cases without much complexity.

To handle the large amount of data and factors, a good solution is decentralized RL. Each agent learns a reduced local problem, decreasing all possible combinations and increasing the number of supported scenarios. Since each agent makes an autonomous decision and computes its own local observation, the model has a higher training diversity with reduced complexity to achieve self-management.
The authors of \cite{Zhao2020} and \cite{Qin2021} proposed a Multi-Agent DRL (MADRL) model with an additional target neural network to stabilize it.  
Authors in \cite{Zhou2022bis} proposed a hybrid solution to optimize user association and MAP trajectory with static clustered users. With pre-deployed MAPs on user clusters, the proposed decentralized DRL is more likely to converge. However, the movement and distribution to the cluster centers remain centralized and must be recomputed after large changes or for each new scenario. Our approach will exploit distributed DRL to handle time-varying variables and free itself from centralized clustering by including the deployment phase. In this case, agents must learn to cooperate to achieve near-optimal solutions. The distributed cooperation will be achieved through the attention mechanism. It allows the model to build its own representation of the input data to handle non-stationary scenarios. It creates a comprehensive context to encourage cooperation and transmit complex messages\cite{Das2018}.

In this paper, we formulate the placement problem to manage MAPs over time. Our solution tackles time-varying user traffic and user mobility through a multi-agent DRL. To achieve real-time behavior, the proposed solution learns to perform a distributed MAP-user positions assignment and schedules the path of MAPs among all users without centralized clustering feedback. Our solution exploits a dual-attention multi-agent DRL model via proximal policy optimization to dynamically move MAPs in 3D. The dual-attention takes into account both users and MAPs information. The cooperation mechanism of our solution allows to manage different scenarios, without a priory information and without re-training, which greatly reduces the complexity.
The paper is organized as follows. Section II presents the system model and formulates the addressed problem. Then, Section III describes our proposed solution, whereas Section IV provides our numerical results. Finally, Section V concludes the paper.

\section{System Model \& Problem Formulation}
\subsection{System Model}
We consider a downlink network composed of $M$ MAPs operating at mmWave frequency and a grounded sub-6GHz Macro-Base-Station (MBS), jointly providing services to $K(t)$ UEs at time $t$. Let $\mathcal{U}(t)=\{1,\dots,K(t)\}$ denote the set of user equipments (UE) and  $\mathcal{M}=\{1,\dots,M\}$ the set of MAPs.
In our system model, we assume that each UE is equipped with two antennas and can communicate either with the sub-6GHz MBS or a mmWave MAP. We assume that each MAP coverage range is determined by its antenna aperture angle $\vartheta_i$. We assume that each UE is associated with the Access Point (AP) providing the maximum signal to noise ratio (called max-SNR algorithm). In addition, we assume that the backhaul network interconnecting the MAPs with the core network is fully provisioned (\ie has sufficient capacity). Thus, we do not optimize backhaul links. However, even with such assumptions, MAP trajectory optimization, which is the focus of this paper, is a crucial task to improve the spectral efficiency of the network by dynamically adapting the location of MAPs \wrt grounded UEs dynamics while limiting interference. In this work, we aim at maximizing total network sum-rate. Let $R_{i,j}(t)$ be the rate experienced by UE $j$ communicating with its serving AP $i$, which is given by the Shannon capacity. 
\begin{align}
    R_{i,j}(t) = B \mathrm{log}_2(1+\mathrm{SINR}_{i,j}(t)).
\end{align}
Here, $B$ is the total system bandwidth and $\mathrm{SINR}_{i,j}(t)$ is the signal-to-interference-plus-noise ratio experienced by the link $i\rightarrow j$, which includes intra-cell and inter-cell interference. In particular, the SINR is affected by the channel path losses, which in turn vary according to several factors, including the 3D location $\ell_i(t)$ of MAP $i$ and $\ell_j(t)$ of UE $j$.

\begin{figure}[!t]
\centering
\includegraphics[width=0.9\columnwidth]{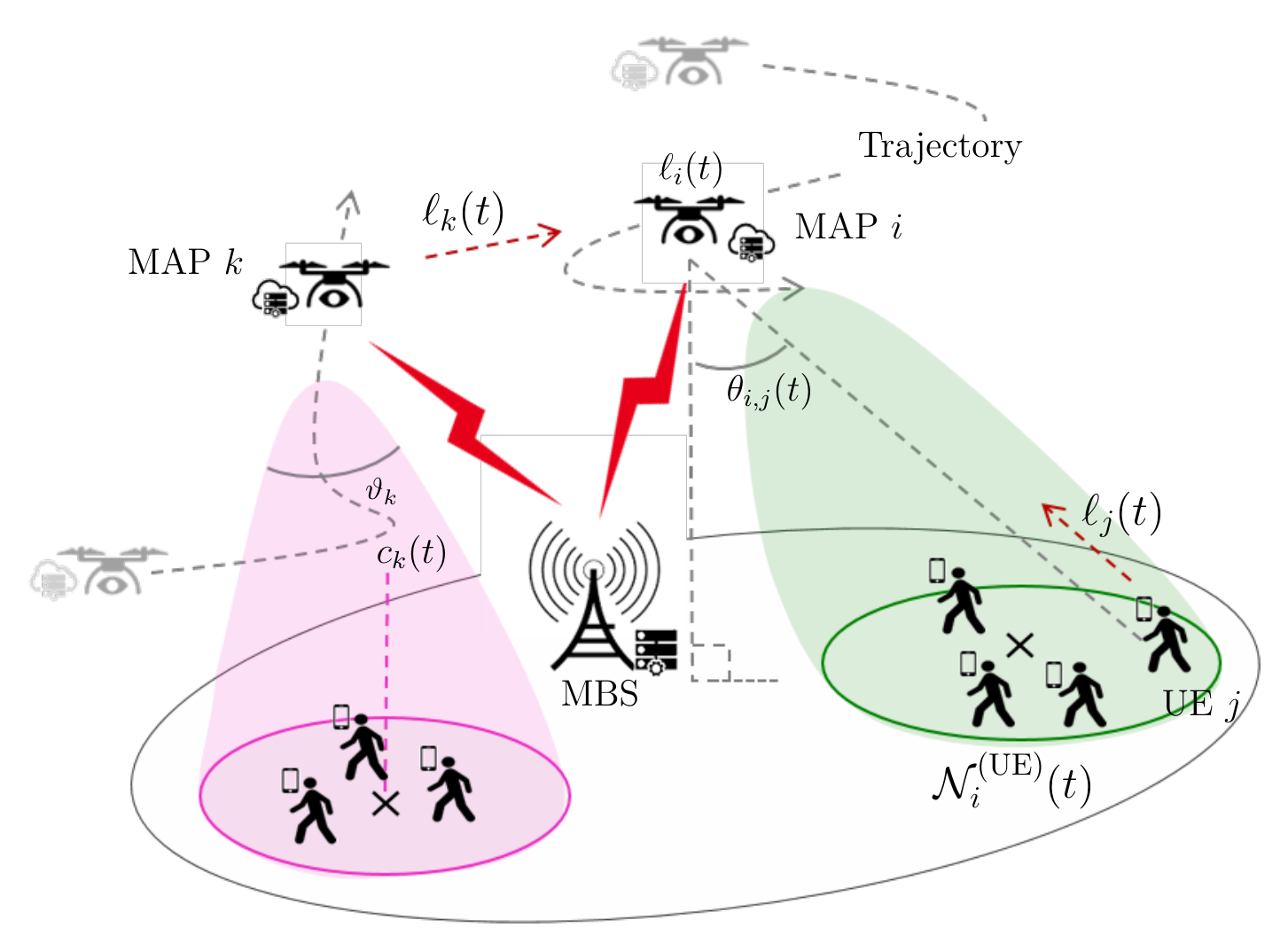}
\vspace{-0.3cm}
\caption{System model with 2 moving MAPs, a fixed MBS and 7 UEs. The MAP $i$ is receiving message from UE $j$ and MAP $k$.}
\vspace{-0.3cm}
\label{fig:system-model}
\end{figure}

We distinguish between ground-to-ground sub-6GHz path loss and air-to-ground mmWave path loss. 
The air-to-ground mmWave path loss depends on Line-of-Sight (LoS) conditions and the euclidean distance $d_{i,j}(t)=\|\ell_i(t) - \ell_j(t)\|$ between MAP $i$ and UE $j$ at time $t$ \cite{Al-Hourani2014}.
\begin{equation}\label{pathloss1}
    {\rm PL}_{i,j}^{(T)}(t) = p(t){\rm PL}_{i,j}^{\rm (Los)}(t) + (1-p(t)){\rm PL}_{i,j}^{\rm (NLos)}(t),
\end{equation}
where $p$ is the LoS probability. Here, ${\rm PL}_{i,j}^{\rm (Los)}$, and ${\rm PL}_{i,j}^{\rm (NLos)}$ are the  LoS and NLoS path loss respectively, given by: 
\begin{equation}
    {\rm PL}_{i,j}^{(l)}(t) = 20\logg{(\frac{4\pi f_c d_{i,j}(t)}{c})} + \chi_{\sigma_l}, ~ l\in\{\rm Los, NLos\}
\end{equation}
where $f_c$ is the carrier frequency, $c$ is the light's velocity, $\chi_{\sigma_l}$ captures the large scale shadowing effect with a standard deviation $\sigma_l$. The LoS probability depends on the relative elevation angle $\theta_{i,j}(t)$ (in radians) between MAP $i$ and UE $j$ (as shown in Figure \ref{fig:system-model}):
\begin{equation}
    p(t) = \frac{1}{1+\alpha e^{-\beta(\theta_{i,j}(t)-\alpha)}}.
\end{equation}
Here, parameters $\alpha$ and $\beta$ depend on the radio environment (e.g. buildings' height and Los condition) and are given in \cite{Al-Hourani2014}.

The ground-to-ground sub-6GHz path-loss \cite{Sun2016} is given as:
\begin{equation}\label{pathloss_ground_ground}
    {\rm PL}_{i,j}^{(T)}(t) = 10\alpha \logg_{10}(d_{i,j}(t)) + \beta + 10\eta \logg_{10}(f_c) + \chi_{\sigma_l},
\end{equation}
where $\alpha$, $\beta$ and $\eta$ depends on radio environment \cite{Sun2016} and Los condition, $f_c$ is the carrier frequency and $\chi_{\sigma_l}$ captures the shadowing effect.
The details of path loss parameters are given in Table \ref{table12}.

\subsection{Problem Formulation}

Let $D_j(t)$ refer to the time-varying data demand of UE $j$ at time $t$. We use $x_{i,j}(t)$ to indicate whether UE $j$ is associated with AP $i$ at time $t$, in which case $x_{i,j}(t)=1$, otherwise $x_{i,j}(t)=0$. In particular, since we assume max-SNR policy for user association, $x_{i,j}(t)$ depends on the location $\ell_i(t)$ of MAP $i$ \wrt UE $j$ location $\ell_j(t)$. Thus, the effective network sum-rate $R(t)$ reads as:
\begin{equation}\label{sumrate}
    R(t) = \sum_{i \in \mathcal{M}} \sum_{j \in \mathcal{U}(t)} x_{i,j}(t)\min\left(D_j(t),R_{i,j}(t)\right).
\end{equation}
We now formulate the following problem to maximize the long-term sum-rate subject to instantaneous constraints:

\begin{align}
    \underset{\mathrmbold{\Psi}(t)}{\mathrm{max}}~~&\lim_{T\rightarrow+\infty}\frac{1}{T}\sum_{t=1}^T \mathbb{E}[R(t)], \tag{$\mathcal{P}_1$} \label{eq:P}\\
    \mathrm{s.t.~}~ & {} x_{i,j}(t) \in \{0,1\}, & \forall i\in \mathcal{M}, j\in \mathcal{U}(t), \tag{$\mathcal{C}_1$} \label{eq:C1}\\
    &\sum _{i\in \mathcal{M}} x_{i,j}(t) = 1, &\forall j \in \mathcal{U}(t), \tag{$\mathcal{C}_2$} \label{eq:C2}\\
    &\sum _{j\in \mathcal{U}(t)} x_{i,j}(t) \leq P, & \forall i \in \mathcal{M}, \tag{$\mathcal{C}_3$} \label{eq:C3}\\
    &\ell_{i}(t) \in \mathcal{L}\subset\mathbb{R}^{3},& \forall i\in\mathcal{M}, \tag{$\mathcal{C}_4$} \label{eq:C4}\\
    &\|\ell_i(t+1) - \ell_i(t)\| \leq \Delta\ell, & \forall i \in \mathcal{M}, \tag{$\mathcal{C}_5$} \label{eq:C5}
\end{align}
where $\mathrmbold{\Psi}(t) = \{\ell_{i}(t), \forall i\}$ and the expectation in \eqref{eq:P} is taken w.r.t the random traffic requests and channels realization, whose statistics are unknown. For the MAPs mobility management problem \eqref{eq:P}, the constraint \eqref{eq:C1} defines $x_{i,j}$ as a binary variable. Eq. \eqref{eq:C2} guarantees that each UE is associated with exactly one AP. If a UE is not connected to a MAP, a connection to the MBS is guaranteed. Constraint \eqref{eq:C3} ensures that every MAP cannot accept more than $P$ connections simultaneously. 
Constraint \eqref{eq:C4} restricts MAPs mobility within an operation zone defined by $\mathcal{L}\subset \mathbb{R}^3$. This restricted zone can for \eg define the maximum and minimum flying altitude.
Finally, constraint \eqref{eq:C5} ensures that a MAP cannot move more than a distance $\Delta \ell$ at a time. In particular, $\Delta \ell$ can be fixed based on the maximum authorized MAP's flight speed. 
Efficiently solving Problem \eqref{eq:P} is challenging. Indeed, the optimal solution of this non-convex combinatorial problem strongly depends on UEs mobility, the dynamics of UEs traffic demands and channel variations. As we jointly optimize the trajectories of multiple MAPs, solutions based on a centralized exhaustive search are unfeasible in practice. To solve this issue, we propose a model-free approach based on MADRL.

\section{Proposed Solution via distributed MADRL}

Our proposed solution is based on MADRL and models each MAP as an RL agent, which learns to make autonomous decisions based only on local observations and some messages received from its neighboring UEs. Agents first autonomously assign themselves a set (cluster) of UEs based on a common ground obtained by a message passing between the MAPs. Then, each agent learns its optimal trajectory, successively deciding its optimal location over time. To do so, at each instant, an agent can make an action from a predefined set $\mathcal{A} = \{\text{forward, backward, up, down, left, right, hover}\}$, corresponding to a movement along the xyz-axis, with a fixed step-size $\Delta\ell$. Our proposed solution is distributed and specifically addresses three challenges: i) a model-free approach, which does not require \emph{a priori} information about the radio environment, channel statistics and UE data demands; ii) efficient representation of agents state observations and design of reward signals to effectively establish a \emph{common ground} between agents iii) flexibility to support size-varying networks, including changes in topology, number and positions of UEs.

\subsection{Background on MADRL}

In a fully observable environment, the decision making process of an agent $i$ can be formalized as a Markov Decision Process (MDP). Formally, a MDP can be defined as a tuple $(\mathcal{S}, \mathcal{A}, \mathcal{T}, \mathcal{R})$ in which $\mathcal{S}$ is the true state space, $\mathcal{A}$ denotes the action space, $\mathcal{T}(s_i(t), a_i(t), s_i(t+1)) = P(s_i(t+1)\given s_i(t), a_i(t))$ is the probability of transitioning to state $s_i(t+1)$ after making action $a_i(t)$ in state $s_i(t)$, which results in an immediate reward $r_i(t) = \mathcal{R}(s_i(t), a_i(t))$. The problem for agent $i$ in a MDP is to find an optimal policy $\pi_i(t): \mathcal{S} \rightarrow \mathcal{A}$ that maximizes the expected sum of perceived ($\gamma$-discounted) rewards $\mathbb{E}_{\pi}[\sum_{\tau=t}^{T_e} \gamma^{\tau-t}r_{i}(\tau)]$ over a time horizon $T_e$, where $\gamma\in[0,1)$. In MADRL, such policy is modeled as a neural network (NN) and is learned by the interaction of several agents with a shared environment. Major challenges appear in this context: the non stationarity of the environment due to the simultaneous interactions of agents. In addition, in our work, an agent has access to only a partial observation of its true state, which is either unknown or difficult to obtain through computation or signals.
To efficiently represent agent states with limited complexity, we propose a novel approach based on neural \emph{attention} mechanism \cite{Vaswani2017}.

\subsection{Dual-Attention Mechanism for Effective Representation Learning}

\begin{figure}[!t]
\centering
\includegraphics[width=\columnwidth]{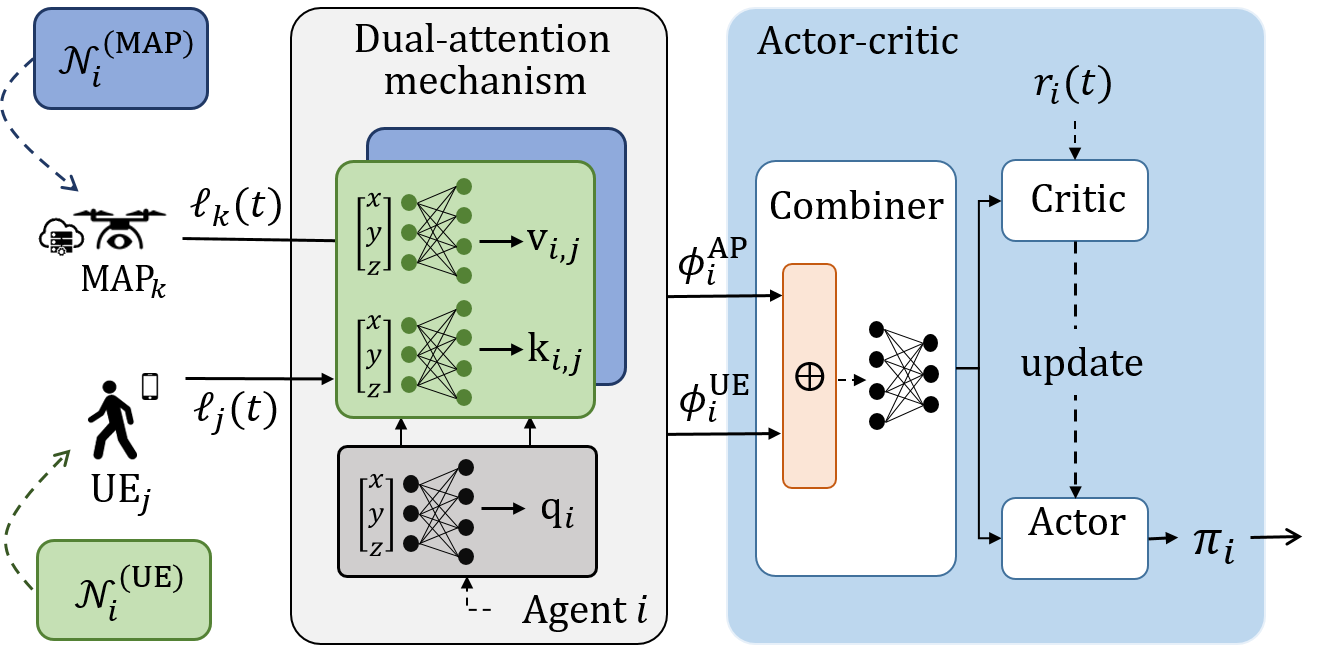}
\caption{Dual-Attention-PPO architecture. Agent $i$ collects message from its neighborhood $\mathcal{N}_i^{(r)}$.}
\vspace{-0.3cm}
\label{fig:architecture}
\end{figure}

To better represent the true states of the agents, we allow information exchange between communicating entities. Specifically, MAP $i$ can collect information about the locations of neighboring MAPs (defined as $\mathcal{N}_i^{(\rm MAP)}(t)$) and neighboring UEs (defined as $\mathcal{N}_i^{(\rm UE)}(t)$), where $\card{\left(\mathcal{N}_i^{(\rm MAP)}(t)\right)} < M$ and  $\card{\left(\mathcal{N}_i^{(\rm UE)}(t)\right)} < K$ to limit the complexity of the information exchange. Then, the agent $i$ learns its actual state representation by encoding the received messages using a \emph{message encoder}. However, in this dynamic environment where the number and position of UEs may change over time, the size of $\mathcal{N}_i^{(r)}(t), \forall r \in \{\rm UE, MAP\}$ and the order of messages may vary accordingly. In this context, the architecture of the \emph{message encoder} must not only be invariant to the varying size of the neighborhood, but also to the permutations of the observed messages. To solve this problem, we adopt the idea from neural \emph{attention} mechanism \cite{Vaswani2017} (see Figure \ref{fig:architecture}). Specifically, $\forall j \in \mathcal{N}_i^{(r)}(t)$, agent $i$ defines $\mathrmbold{k}_{i,j}^{(r)}(t) = \params_{i,\rm k}^{(r)} (\ell_i(t) - \ell_j(t))^T \in \mathbb{R}^n$, $\mathrmbold{v}_{i,j}^{(r)}(t) = \params_{i,\rm v}^{(r)} (\ell_i(t) - \ell_j(t))^T \in \mathbb{R}^n$, referred to as the relative \emph{key} and \emph{value} associated with message of entity $j$. In addition, agent $i$ computes its \emph{query} vector $\mathrmbold{q}_{i}^{(r)}(t) = \params_{i, \rm q}^{(r)} \ell_i(t)^T \in \mathbb{R}^n$. Here, the encoding matrices $\params_{i, \rm k},  \params_{i, \rm v}, \params_{i, \rm q} \in \mathbb{R}^{n\times 3}$ are learnable parameters (\egn, a hidden layer of dimension $n$).
Then, the \emph{value} of the messages are aggregated independently to compute agent $i$ state representation $\phi_i^{(r)} \in \mathbb{R}^n$:
\begin{equation}
    \phi_i^{(r)}(t) = \sum_{j\in\mathcal{N}_i^{(r)}(t)}\alpha_{i,j}(t) \mathrmbold{v}_{i,j}^{(r)}(t), 
\end{equation}
where $\alpha_{i,j}(t)$ defines the interaction vector of agent $i$ \wrt entity $j$ and reads as follows:
\begin{equation}
    \alpha_{i,j}(t) = \mathrm{softmax}\left(\left[ \frac{\mathrmbold{q}_{i}^{(r)} {\mathrmbold{k}_{i,p}^{(r)}}^{T}}{\sqrt{n}} \right]_{p\in\mathcal{N}_i^{(r)}(t)}\right)_j.
\end{equation}
Here, $\mathrm{softmax}$ is the normalized exponential function. By construction, the size of $\phi_i^{(r)}$ is constant and equal to $n$ and does not vary with the size of the neighborhood $\mathcal{N}_i^{(r)}(t)$ as desired. In our solution, $\phi_i^{(\rm MAP)}$ captures the relative perception of MAP $i$ \wrt to others neighboring MAPs. In contrast, $\phi_i^{(\rm UE)}$ captures the relative perception of MAP $i$ \wrt to neighboring UEs. Next, as shown in Figure \ref{fig:architecture}, we combine these two dual representations to serve as input to an \emph{actor-critic} framework, which we optimize in an end-to-end manner using the well-known proximal policy optimization (PPO). Specifically, we minimize the $(\epsilon_1, \epsilon_2)$-clipped proximal loss proposed in our previous work \cite{Sana2021}.

\subsection{Designing Effective Reward Function}

To effectively learn a common ground between agent, we adopt a hierarchical approach for designing the reward signal. Our goal is first to maximize user coverage, then maximize network sum-rate. To do so, during the training phase, each agent $i$ learns to maximize the following reward function:
\begin{align}\label{eq:reward}
    r_i(t) = (\delta_i(t)-1)d_{i}(t) + \delta_i(t) (R(t)-d_0).
\end{align} 
Here, $\delta_i(t) = \mathds{1}(d_i(t) \leq d_0)$, where $d_0$ is a reference distance and $d_i(t) = \|\ell_i(t) - \ell_i^*(t)\|$ is the distance of MAP $i$ to its optimal location $\ell_i^*(t)$ (\wrt \eqref{eq:P}). Since this location is not known a priori, we approximate it during the training phase with the location of the nearest assigned centroid $c_i(t)$ obtained after clustering UEs using \eg Kmeans algorithm (Algorithm \ref{alg:training}). In particular, we obtain the $z$-coordinate of centroid by computing the minimal altitude at which a MAP with an aperture angle of $\vartheta_i$ would cover the UE cluster. In this way, we push agent $i$ to first maximize user coverage (first term in Equation \ref{eq:reward}) and then network sum-rate (second term in Equation \ref{eq:reward}). It is worth noting that the clustering operation only serves during the training phase for learning a \emph{common ground}. In contrast, no central coordinator is required during testing and MAPs autonomously coordinate themselves to optimize \eqref{eq:P} (Algorithm \ref{alg:testing}). 

\begin{remark}
In practice, we normalize $d_i(t)$ by a predefined maximum distance and $R(t)$ by to its average value (\wrt environment randomness). Also, instead of the step function $\delta_i(t)$, we use a smooth Gompertz function (see Figure \ref{fig:reward2}), a generalized logistic function $\delta_i(t)= 1 - 0.9 e^{-e^{-0.06(d_i(t) - d_0)}}$.
\end{remark}

\begin{algorithm}[!t]
\small
\caption{Training algorithm}\label{alg:training}
\KwIn{
Init model weights $w$, environment state $\mathcal{S}$
}
\For {$run=1...\mathrm{run}$}{
Randomly deploy agents in the cell\\
Agents receive UE centroids location $c_i(0)$\\
Agents compute their neighborhood $\mathcal{N}_i^{(r)}(0)$\\
Agents gather messages from $\mathcal{N}_i^{(r)}(0)$\\
Agents computes their state representation $\phi_i^{(r)}(0)$\\
\For{$t=1 \dots T_e$}{
Agents select and execute $a_i(t)$\\
Agents receive their nearest assigned centroid $c_i(t)$\\
Agents receive rewards $r_i(t)$\\
Update the environment state $\mathcal{S}$\\
Agents compute their neighborhood $\mathcal{N}_i^{(r)}(t)$\\
Agents gather messages from $\mathcal{N}_i^{(r)}(t)$\\
Agents compute their state representation $\phi_i^{(r)}(t)$\\
}
Compute $(\epsilon_1, \epsilon_2)$-clipped proximal loss\\
Performs gradient descent step with Adam\\
Update policies $\pi_{i}, \forall i\in \mathcal{M}$\\
}
\KwOut{policies $\pi_{i}$} 
\end{algorithm}

\begin{algorithm}[!t]
\small
\caption{Testing algorithm}\label{alg:testing}
\KwIn{
Load model weights $w$.\\
Trained agent policies $\pi_{i}$
}
\For {$run=1...\mathrm{run}$}{
\For{$t=1 \dots T_e$}{
Agents computes their neighborhood $\mathcal{N}_i^{(r)}(t)$\\
Agents gather messages from $\mathcal{N}_i^{(r)}(t)$\\
Agents computes their state representation $\phi_i^{(r)}(t)$\\
Agents select and execute $a_i(t)$\\
Update the environment state $\mathcal{S}$\\
Compute $R(t)$
}
}
\KwOut{E[$R(t)$]} 
\end{algorithm}

\section{Numerical Results}

We perform the training of our proposed MADRL algorithm for $10000$ Monte-Carlo \emph{runs}. For each run, we deploy $M=3$ MAPs moving with a step size $\Delta\ell=5\m$; we also deploy $K=25$ UEs in $M$ centroids of radius $25\m$, randomly sampled in a $200\m$ by $200\m$ area. During the training phase, the UEs are static for $T_e=300$ iterations while they follow a random way-point centroid mobility at $0.8\mathrm{m/s}$ during the testing phase. As the model has not been trained with specific mobility model, it is able to support time-varying constraints. 
We set the agent maximum neighborhood size $\max{\left( \card{\left(\mathcal{N}_i^{(\rm UE)}(t)\right)}\right)}=15$ and $\max{\left(\card{\left(\mathcal{N}_i^{(\rm MAP)}(t)\right)}\right)}=3$ composed of nearest entities.
The UEs traffic follows a Poisson distribution of parameter $k=1000 \Mbps$. We consider a Nakagami fast-fading model of parameter $\nu=1$ for each channel and other channel parameters are given in Table \ref{table12}.
We train each model with a learning rate equals to $1e-4$ and $\gamma=0.6$. We set PPO-clips $(\epsilon_1, \epsilon_2) = (0.01,0.5)$, and compose the message encoders with one multi-layer perceptron (MLP) of $n=128$ neurons. The actor and critic comprises also one MLP of $2n$ neurons. 

\begin{table}[!htp]
\caption{Simulation Parameters}
    \centering
    \label{table12}
    \scalebox{\tablescale}{
    \begin{tabular}{l||c|c }
    \hline
    \textbf{Channel Parameters} & \textbf{MBS} & \textbf{MAP} \\
      \hline
      Carrier Frequency $f_{c}$ & $2\GHz$ & $28\GHz$\\
      \hline
      Bandwidth $B$ & $10\MHz$ & $500\MHz$ \\
      \hline
      Thermal Noise $N_{0}$ & \multicolumn{2}{c}{$-174\dBm/\Hz$} \\
      \hline
      Shadowing Variance $\sigma_l^2$ & $3\dB$ & $12\dB$\\
      \hline
      Antenna Gain & $17~\dBi$ & Directive \cite{Sana2021} \\
      \hline
      Antenna Aperture Angle & $180°$ & $90°$ \\
      \hline
      LoS Path Loss Parameter & $\alpha=2$ & $\alpha=10.37$ \\ & $\beta=31.4$ & $\beta=0.05$ \\ & $\eta=2.1$ & \\
      \hline
      NLoS Path Loss Parameter & $\alpha=3.5$ & $\alpha=35.85$ \\ & $\beta=24.4$ & $\beta=0.04$ \\ & $\eta=1.9$ & \\
    \end{tabular}}
    \label{tab:channel}
    \vspace{-0.3cm}
\end{table}

\noindent
\textbf{Benchmarks.} We define two benchmark solutions. 
The first benchmark (referred to as \textbf{Centralized Ben.}) pre-computes the centroids of UE clusters using a centralized Kmeans algorithm. Then, each centroid is assigned to the closest MAP, whose trajectory is planned using Dijkstra's algorithm. The second benchmark (referred to as \textbf{SA-PPO})  is similar to our proposed solution, in which we employ a single attention mechanism \wrt UEs without any cooperation between MAPs. 

\begin{figure}[!htp]
\centering
\vspace{-0.3cm}
\includegraphics[width=\columnwidth]{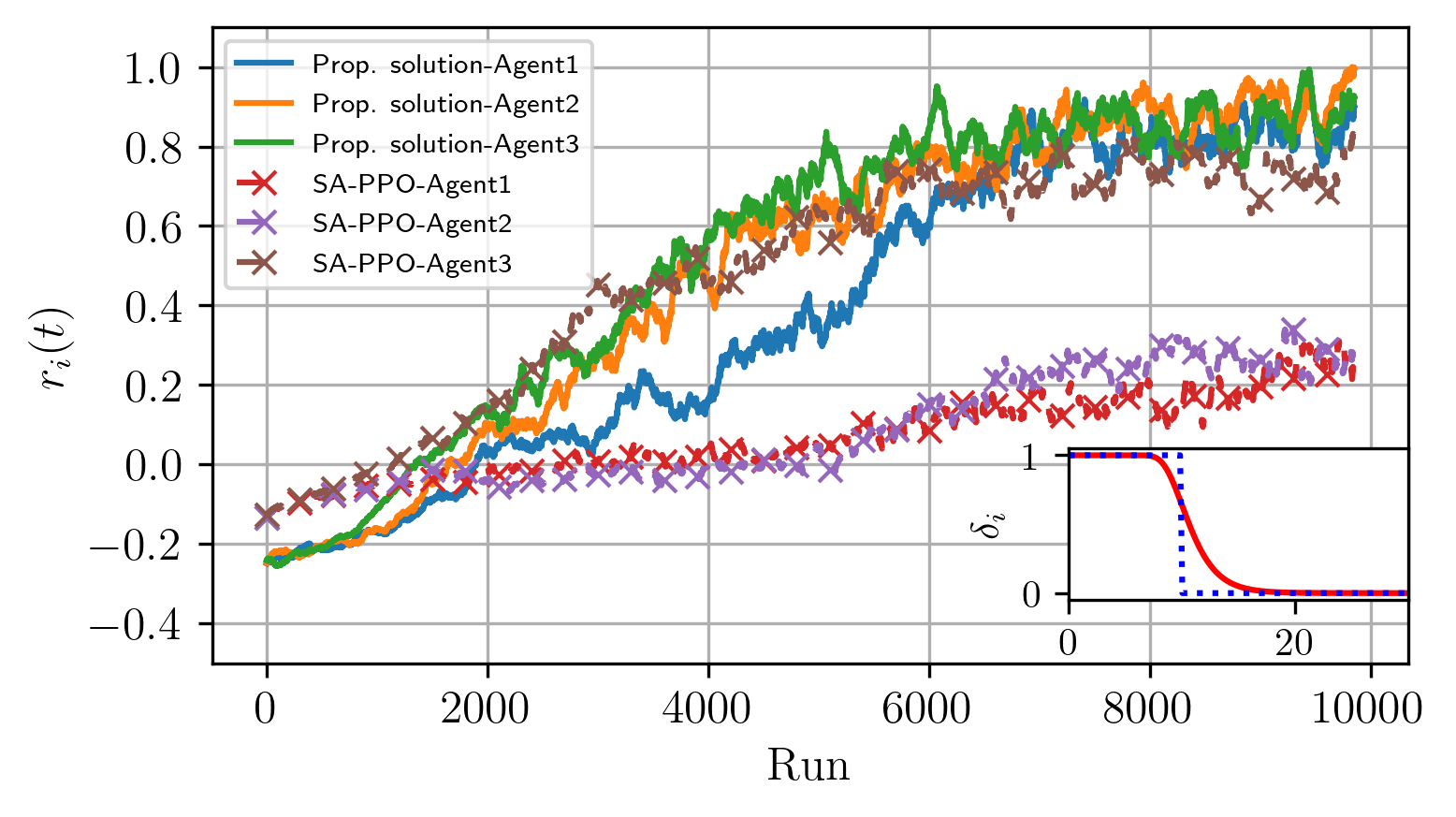}
\vspace{-0.6cm}
\caption{Proposed solution and single-attention training rewards $r_i(t)$ for each agent and associated step and Gompertz functions $\delta_i(\cdot)$.} 
\label{fig:reward2}
\end{figure}

\noindent
\textbf{Learning convergence}. We first compare the rewards of SA-PPO and our proposed solution. As shown in Figure \ref{fig:reward2}, both models converge for the complex \eqref{eq:P} problem. Indeed, each method obtains a positive reward for all agents, which means that each agent achieves a placement at a distance less than $d_0$ to the centroid center (Equation \ref{eq:reward}). Especially, our solution ends with a higher reward for each agent compared to the SA-PPO training. The difference comes from the lack of explicit cooperation among 
SA-PPO agents, 
resulting in a lower reward.  Indeed, if two MAPs end up serving the same set of UEs, the sum-rate, which is a global performance, drops sharply. In contrast, by using our proposed dual-attention mechanism, agents converge to the same behavior and cooperatively distribute MAPs to UE clusters.

\begin{figure}[!htp]
\centering
\vspace{-0.3cm}
\includegraphics[width=0.9\columnwidth]{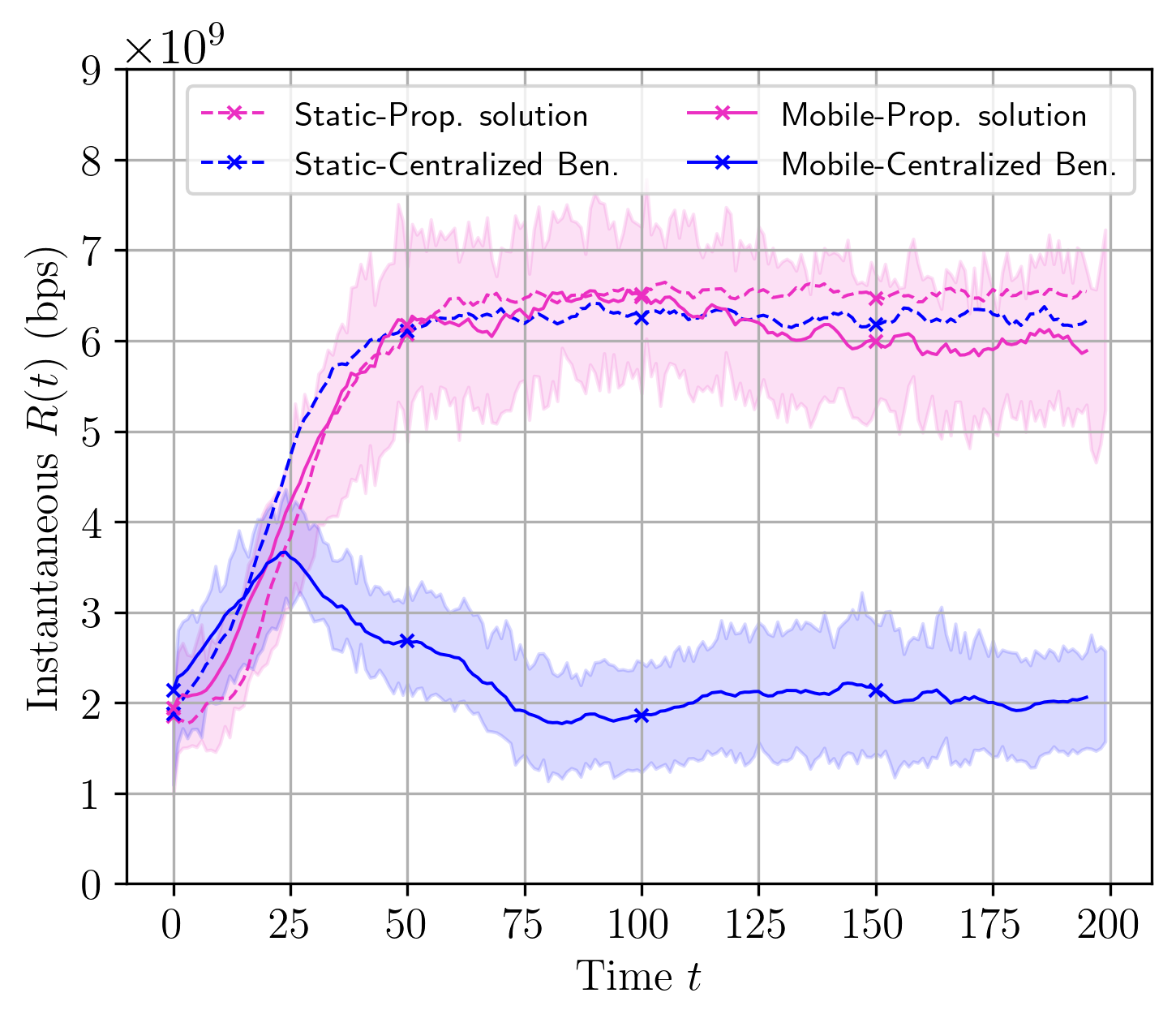}
\vspace{-0.3cm}
\caption{Instantaneous sum-rate comparison between our solution and the benchmark for static and mobile UEs over time. Results are averaged over $50$ random deployments.}
\label{fig:static_200}
\end{figure}

\noindent
\textbf{Distributed cooperation}. In this step, we validate that our solution is able to deploy and place agents with \textit{Static} and \textit{Moving} UEs thanks to cooperation and we compare our solution with the centralized benchmark for a $\tau_c=200$. UE traffic requests are always dynamic and moving MAPs introduce fluctuating level of interference. For the \textit{Static} scenario, Figure \ref{fig:static_200} shows that our model matches the sum-rate performance of the centralized approach. 
Therefore, our solution achieves a fast location negotiation explained by the 10 iterations latency on the deployment phase. The first iterations are used to exchange messages to discover and allocate agents to the negotiated positions. Furthermore, it proves that our model is able to compute and focus a virtual point via message exchange, which leads to a distributed clustering behavior.
Concerning the \textit{Moving} scenario, our solution renegotiates its positions to follow network dynamics robustly. Figure \ref{fig:static_200} shows the importance of not considering only the MAP deployment phase, due to the change of network during MAPs time of flight, preventing a performance drop. It is now important to extend the time window to ensure that our model does not also suffer from a performance drop.

\begin{figure}[!htp]
\centering
\includegraphics[width=\columnwidth]{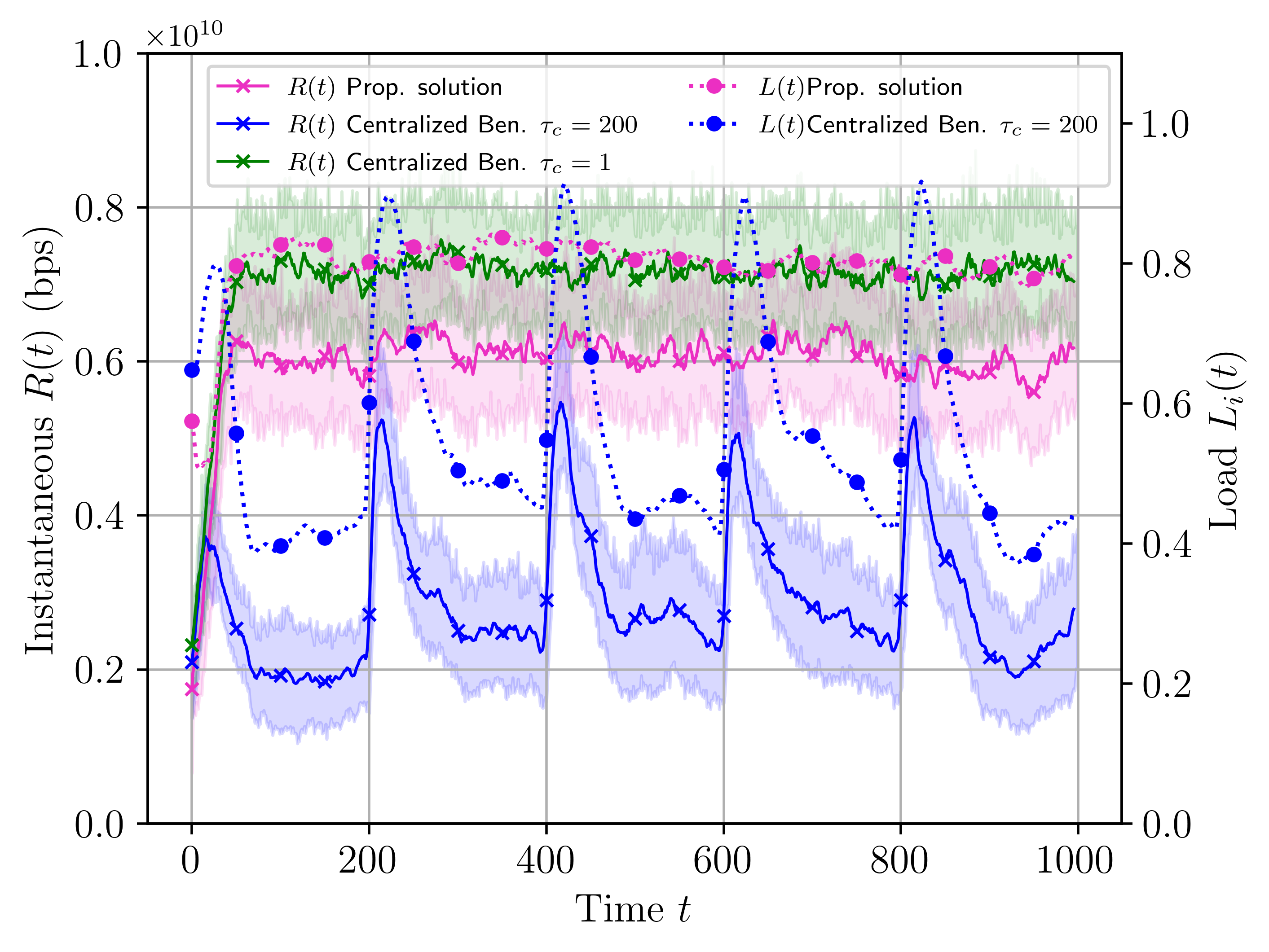}
\vspace{-0.6cm}
\caption{Instantaneous network sum-rate evolution over time with $\tau_c=\{1,200\}$. Results are averaged over $50$ random deployments.}
\label{fig:moving_1000}
\end{figure}

\noindent
\textbf{Robustness to dynamic networks}. In this step, we investigate how the centralized benchmark and our solution support highly dynamic networks with or without centralized user clustering feedback for a long time period ($t=[0,1000]$). Figure \ref{fig:moving_1000} shows that the centralized benchmark is attractive when the feedback on clustering occurs every time slot ($\tau_c=1$). This comes with a very high complexity. When the feedback becomes less frequent ($\tau_c=200$), the benchmark is not robust to network dynamics and the performance drops after $30$ time slots. This is due to the fact that the clustering information is outdated because of major changes in the network.
Figure \ref{fig:moving_1000} shows that our solution is able to guarantee a sum-rate and load level without any drop when the entire network configuration changes several times and then achieve real-time behavior. 
Dual-attention agents trade $20\%$ of the network \emph{load} $L_i(t)$, representing the proportion of users connected to a MAP, against outlier UEs with isolated behavior, with a guaranteed expected sum-rate. 
At this time scale, the few iterations used to the beginning of the negotiation become negligible and this scenario demonstrates the importance of the parameter $\tau_c$.

\begin{figure}[!htp]
\centering
\vspace{-0.3cm}
\includegraphics[width=\columnwidth]{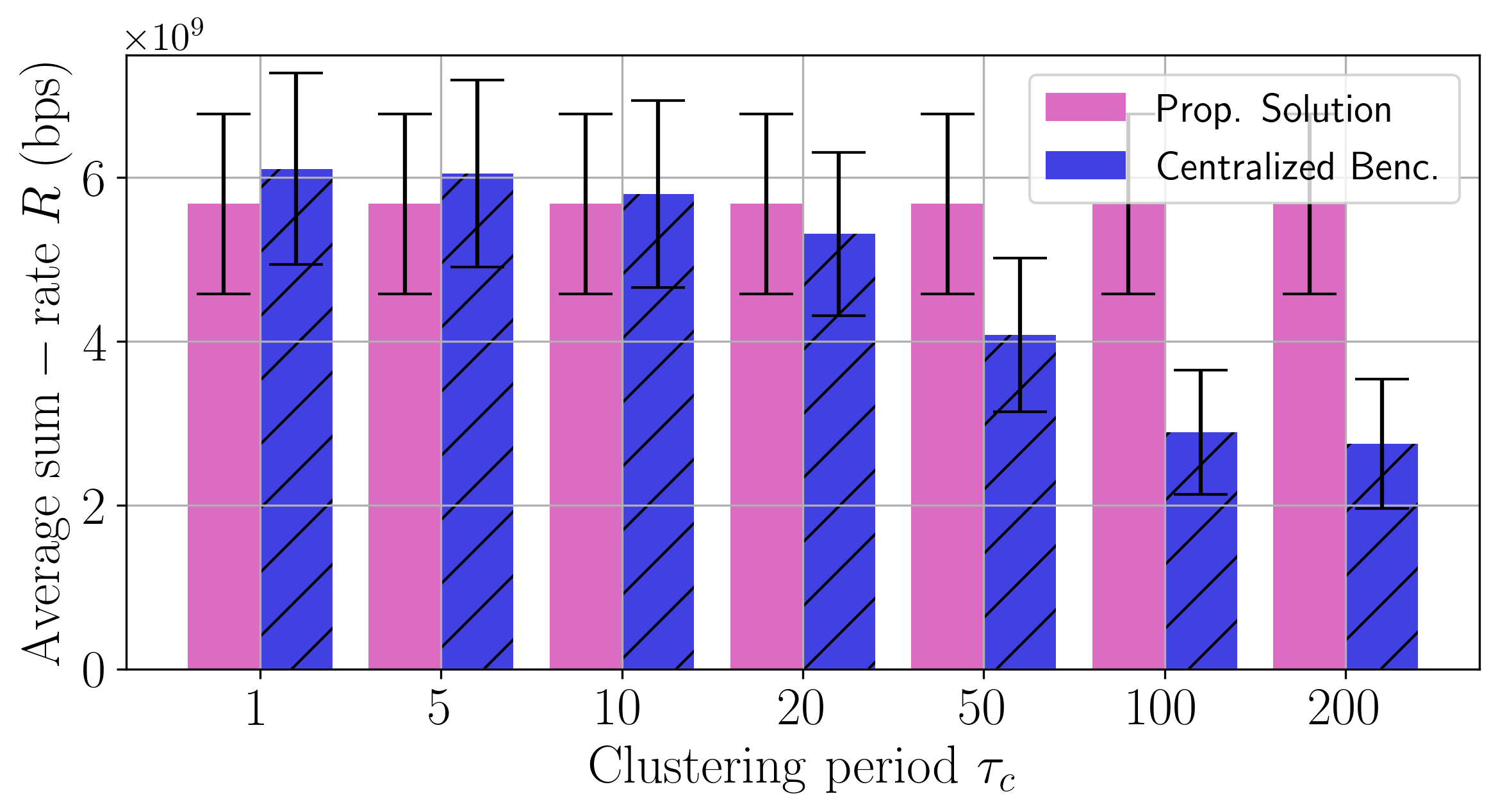}
\vspace{-0.6cm}
\caption{Average sum-rate $R$ for our solution and the centralized benchmark for different $\tau_c$ clustering periods.}
\label{fig:hist}
\end{figure}

\noindent
\textbf{Trade-off on clustering}. 
Figure \ref{fig:hist} compares our solution with the centralized benchmark for different clustering periods $\tau_c$. The benchmark performance strongly depends on $\tau_c$ which defines the age of information. In contrast, our solution does not depend on clustering updates coming from a central coordinator. The centralized approach must cluster the UEs every $\tau_c=10$ time slots to match our solution. 

\section{Conclusion}
In this paper, we study the optimization of multi-MAP 3D trajectory for dynamic 5G networks. To this end, we addressed the mobility management of MAPs under time-varying user traffic, user mobility and interference. We proposed a dual-attention MADRL solution capable of self-managing a flexible infrastructure using cooperative MAPs. The proposed solution learns to perform distributed assignment of MAP-user positions and schedules the MAP path among all users without centralized user clustering feedback. Agents converge to the same behavior and cooperatively distribute MAPs to UE clusters. The cooperation mechanism also allows to manage different scenarios, without a priory information and without re-training, which significantly reduces complexity. Our solution does not depend on clustering updates coming from a central coordinator and is robust to network dynamics.
In future work, the solution will be extended to handle more dynamic parameters and imperfections such as the backhaul connectivity or imperfect beamforming. Moreover, our solution will include additional metrics such as energy and deployment cost.

\section*{Acknowledgment}
This work was supported by the European Union H2020 Project \mbox{DEDICAT 6G} under grant no. 101016499. The contents of this publication are the sole responsibility of the authors and do not in any way reflect the views of the EU.

\bibliographystyle{ieeetr}
\bibliography{./Biblio}

\begin{thebibliography}{10}

\bibitem{Maman2022b}
M.~Maman, E.~Catte, M.~Sana, M.~Girmay, V.~Maglogiannis, D.~Naudts, H.~Lee,
  F.~Carrez, A.~Anttonen, Y.~Fernandez, J.~Moreno, V.~Lamprousi, and
  V.~Stavroulaki, ``{Coverage Extension as a Service for Flexible 6G Networks
  Infrastructure},'' in {\em 2022 IEEE Globecom Workshops (GC Wkshps)},
  pp.~1329--1334, 2022.

\bibitem{Catte2022}
E.~Catté, M.~Sana, and M.~Maman, ``{Cost-Efficient and QoS-Aware User
  Association and 3D Placement of 6G Aerial Mobile Access Points},'' in {\em
  2022 Joint European Conference on Networks and Communications \& 6G Summit
  (EuCNC/6G Summit)}, pp.~357--362, 2022.

\bibitem{Alsharoa2018}
A.~Alsharoa, H.~Ghazzai, M.~Yuksel, A.~Kadri, and A.~E. Kamal, ``{Trajectory
  Optimization for Multiple UAVs Acting as Wireless Relays},'' in {\em 2018
  IEEE International Conference on Communications Workshops (ICC Workshops)},
  pp.~1--6, 2018.

\bibitem{Wu2018}
Q.~Wu, Y.~Zeng, and R.~Zhang, ``{Joint Trajectory and Communication Design for
  Multi-UAV Enabled Wireless Networks},'' {\em IEEE Transactions on Wireless
  Communications}, vol.~17, no.~3, pp.~2109--2121, 2018.

\bibitem{Pan2022}
Y.~Pan, X.~Da, H.~Hu, Y.~Huang, M.~Zhang, K.~Cumanan, and O.~A. Dobre, ``{Joint
  Optimization of Trajectory and Resource Allocation for Time-Constrained
  UAV-Enabled Cognitive Radio Networks},'' {\em IEEE Transactions on Vehicular
  Technology}, vol.~71, no.~5, 2022.

\bibitem{Rozhina2018}
R.~Ghanavi, E.~Kalantari, M.~Sabbaghian, H.~Yanikomeroglu, and A.~Yongacoglu,
  ``{Efficient 3D aerial base station placement considering users mobility by
  reinforcement learning},'' in {\em 2018 IEEE Wireless Communications and
  Networking Conference (WCNC)}, pp.~1--6, 2018.

\bibitem{Liu2019}
X.~Liu, Y.~Liu, and Y.~Chen, ``{Reinforcement Learning in Multiple-UAV
  Networks: Deployment and Movement Design},'' {\em IEEE Transactions on
  Vehicular Technology}, vol.~68, no.~8, pp.~8036--8049, 2019.

\bibitem{Saxena2019}
V.~Saxena, J.~Jaldén, and H.~Klessig, ``{Optimal UAV Base Station Trajectories
  Using Flow-Level Models for Reinforcement Learning},'' {\em IEEE Transactions
  on Cognitive Communications and Networking}, vol.~5, no.~4, pp.~1101--1112,
  2019.

\bibitem{Zhang2021}
W.~Zhang, Q.~Wang, X.~Liu, Y.~Liu, and Y.~Chen, ``{Three-Dimension Trajectory
  Design for Multi-UAV Wireless Network With Deep Reinforcement Learning},''
  {\em IEEE Transactions on Vehicular Technology}, vol.~70, no.~1,
  pp.~600--612, 2021.

\bibitem{Ji2022}
J.~Ji, K.~Zhu, and L.~Cai, ``{Trajectory and Communication Design for
  Cache-Enabled UAVs in Cellular Networks: A Deep Reinforcement Learning
  Approach},'' {\em IEEE Transactions on Mobile Computing}, 2022.

\bibitem{Ding2020}
R.~Ding, F.~Gao, and X.~S. Shen, ``{3D UAV Trajectory Design and Frequency Band
  Allocation for Energy-Efficient and Fair Communication: A Deep Reinforcement
  Learning Approach},'' {\em IEEE Transactions on Wireless Communications},
  vol.~19, no.~12, pp.~7796--7809, 2020.

\bibitem{Zhao2020}
N.~Zhao, Z.~Liu, and Y.~Cheng, ``{Multi-Agent Deep Reinforcement Learning for
  Trajectory Design and Power Allocation in Multi-UAV Networks},'' {\em IEEE
  Access}, vol.~8, pp.~139670--139679, 2020.

\bibitem{Qin2021}
Z.~Qin, Z.~Liu, G.~Han, C.~Lin, L.~Guo, and L.~Xie, ``{Distributed UAV-BSs
  Trajectory Optimization for User-Level Fair Communication Service With
  Multi-Agent Deep Reinforcement Learning},'' {\em IEEE Transactions on
  Vehicular Technology}, vol.~70, no.~12, 2021.

\bibitem{Zhou2022bis}
S.~Zhou, Y.~Cheng, X.~Lei, Q.~Peng, J.~Wang, and S.~Li, ``{Resource Allocation
  in UAV-assisted Networks: A Clustering-Aided Reinforcement Learning
  Approach},'' {\em IEEE Transactions on Vehicular Technology}, pp.~1--16,
  2022.

\bibitem{Das2018}
A.~Das, T.~Gervet, J.~Romoff, D.~Batra, D.~Parikh, M.~G. Rabbat, and J.~Pineau,
  ``Tarmac: Targeted multi-agent communication,'' {\em CoRR},
  vol.~abs/1810.11187, 2018.

\bibitem{Al-Hourani2014}
A.~Al-Hourani, S.~Kandeepan, and S.~Lardner, ``{Optimal LAP Altitude for
  Maximum Coverage},'' {\em IEEE Wireless Communications Letters}, vol.~3,
  no.~6, pp.~569--572, 2014.

\bibitem{Sun2016}
S.~Sun, T.~S. Rappaport, {\em et~al.}, ``{Propagation Path Loss Models for 5G
  Urban Micro- and Macro-Cellular Scenarios},'' in {\em Proc. IEEE Vehicular
  Technology Conference (VTC Spring)}, pp.~1--6, 2016.

\bibitem{Vaswani2017}
A.~Vaswani, N.~Shazeer, N.~Parmar, J.~Uszkoreit, L.~Jones, A.~N. Gomez,
  L.~Kaiser, and I.~Polosukhin, ``Attention is all you need,'' in {\em
  Proceedings of the 31st International Conference on Neural Information
  Processing Systems}, NIPS'17, p.~6000–6010, Curran Associates Inc., 2017.

\bibitem{Sana2021}
M.~Sana, N.~di~Pietro, and E.~{Calvanese Strinati}, ``{Transferable and
  Distributed User Association Policies for 5G and Beyond Networks},'' in {\em
  Proc. IEEE International Symposium on Personal, Indoor and Mobile Radio
  Communications (PIMRC)}, pp.~966--971, 2021.

\end{thebibliography}
\end{document}